# Radio supernovae as TeV gamma-ray sources

J.G. Kirk[1], P. Duffy[1], and Lewis Ball[2]

[1] Max-Planck-Institut für Kernphysik, Postfach 10 39 80, D-69029 Heidelberg, Germany
[2] Research Centre for Theoretical Astrophysics, University of Sydney, N.S.W. 2006, Australia

**Abstract.** When applied to the blast wave formed by the explosion of a massive star as a supernova (SN), the theory of diffusive particle acceleration at shock fronts predicts a very high energy density in cosmic rays. Almost immediately after particles begin to be injected into the process, the cosmic ray pressure rises until comparable to the ram-pressure encountered by the shock front. Those supernovae which are observed in the radio band i.e., radio supernovae (RSNe), provide direct evidence of particle acceleration in the form of synchrotron emitting electrons. Furthermore, these objects are particularly interesting, since they are usually surrounded by a relatively dense confining medium. The acceleration of cosmic rays can then lead to the production of very high energy (VHE) gamma-rays which arise from collisions between energetic particles and target nuclei. We estimate the cosmic ray energy density assuming a fraction $\phi \lesssim 1$ of the energy available at the shock front is converted into cosmic rays. Combining this with the parameters describing the environment of the SN progenitor, as deduced from observations, and from more detailed modelling, we compute the flux at Earth $F(> 1\,\text{TeV})$ of photons of energy above 1 TeV. For the relatively weak but nearby supernova SN1987A we predict $F(> 1\,\text{TeV}) = 2 \times 10^{-13}$ photons s$^{-1}$ cm$^{-2}$ before the shock front encounters the ring of dense matter seen by the Hubble Space Telescope. Subsequently, the flux is expected to rise further. The medium around SN1993J in M81 is thought to have a density profile $\rho \propto r^{-3/2}$, (with $r$ the distance from the point of explosion) for which we predict a roughly constant flux of $F(> 1\,\text{TeV}) = 2 \times 10^{-12}$ photons s$^{-1}$ cm$^{-2}$. Once the shock emerges into the region where $\rho \propto r^{-2}$, which is expected at larger distances, the flux should decrease in proportion to $t^{-1}$. This object thus presents an interesting target for observation by telescopes which detect the Čerenkov light emitted by the air showers from VHE photons. A detection would provide observational confirmation of cosmic ray acceleration in supernovae.

**Key words:** acceleration of particles – shock waves – cosmic rays – gamma rays: observations – SN1987A – SN1993J

## 1. Introduction

The strong shock fronts associated with supernova remnants are currently thought to be responsible for the acceleration of galactic cosmic rays of energy below about $10^{14}$ eV. Apart from indirect arguments based upon the total energy budget required to sustain these particles against loss out of the galaxy, the main support for this hypothesis comes from the existence of a fairly well-developed theory for particle acceleration – diffusive acceleration – which has been successfully applied to supernova shocks by several groups (Drury et al. 1989; Dorfi 1990; Jones & Kang 1990; Berezhko et al. 1994). However, there has not yet been any direct empirical confirmation of the theory. Estimates of the flux of high energy photons and neutrinos from galactic type II supernovae have been made by Berezinskii & Ptuskin (1988). More recently, it has been pointed out that supernova remnants might be observable as sources of TeV photons using the atmospheric Čerenkov technique (Drury et al. 1994). Such an observation would be an important test of the theory, but the identification of a suitable target is not straightforward (Aharonian et al. 1994). In this *Letter*, we show that the theory of diffusive particle acceleration predicts TeV photons not only from *supernova remnants* (SNRs), but also from powerful *radio supernovae* shortly after explosion.

Very high energy gamma-rays are produced by energetic protons (cosmic rays) when they interact with target nuclei, and create neutral pions. Thus, in any given volume of space, the production rate is proportional to the density of cosmic rays and the total mass of target nuclei. Compared to freshly exploded stars, supernova remnants occupy a relatively large volume containing a high target mass (of the order of a few solar masses). However, the pressure (or energy density) of cosmic rays within SNRs is much smaller than can be expected in the early phases of a supernova explosion which is confined inertially by a very dense medium (e.g., the stellar wind of the progenitor). Conveniently, confined explosions also manifest themselves as radio sources as soon as the shock front has penetrated sufficiently far into the surrounding plasma to enable the photons to escape through it. Over a dozen such 'Radio Supernovae' (RSNe) have been observed to date. Of these, two are in galaxies close enough to our own to permit, in principle, detection of the gamma-rays: SN1987A, which is an intrinsically weak source, but very close-by, and SN1993J, a very powerful RSNe situated in M81 at a distance of about 3.6 Mpc.

## 2. Diffusive particle acceleration in SN1987A

Despite the fact that theoretical arguments (Berezinskii & Ptuskin 1988; Völk & Biermann 1988) have emphasised that the early phase of a supernova explosion in an inhomogeneous medium is important for the acceleration of high energy cosmic rays, almost all detailed investigations of diffusive particle acceleration in SNRs have dealt with the case of a star exploding into uniform homogeneous surroundings. Furthermore, attention has been focussed on the nuclear component of the accelerated particles, since it is these which must carry off most of the energy and thus ultimately determine the overall acceleration efficiency. The case of SN1987A changed the situation in both respects. The progenitor star is known to have had a strong wind, which significantly modified the environment, and the theory of diffusive acceleration was applied in an effort to explain the relativistic electrons responsible for the radio emission, rather than the acceleration of cosmic rays (Ball & Kirk 1992, henceforth BK). Although it is possible to construct models which account for the radio emission, the observed spectrum requires a shock of compression ratio about 2.7, much less than the value 4 expected for a strong supernova shock front expanding into the stellar wind of its progenitor. In view of the speed of the shock front ($\approx 20,000\,\mathrm{km\,s^{-1}}$), it is unrealistic to suppose the Mach number could be reduced to the required value (about 2) merely by preheating the upstream medium. Instead, BK suggested that the shock front is modified by cosmic rays, which exert a substantial pressure on the upstream medium. This kind of modified shock front is well-known in the nonlinear theory of diffusive shock acceleration, and can result in an extended upstream precursor and a gas subshock of relatively low compression ratio (e.g., Drury 1983). Following up on this suggestion, 'two-fluid' simulations similar to those of supernova remnants (e.g., Dorfi 1990) were performed for a supernova exploding into an inhomogeneous region. The first density profile to be investigated was

$$\rho(r) = \dot{M}/(4\pi r^2 v_\mathrm{w}) , \qquad (1)$$

corresponding to that of a progenitor undergoing steady mass-loss at a rate $\dot{M}$ via a stellar wind of constant velocity $v_\mathrm{w}$ (Duffy et al. 1993; Kirk et al. 1994a). Once the hydrodynamic structure of the flow was found, the radio emission was computed by considering the acceleration of electrons, which can be treated as test particles because of their negligible contribution to the energy density. More recently, the computations have been extended to a more realistic picture of the surroundings of SN1987A (Duffy et al. 1994 henceforth DBK), which is thought to consist of a freely expanding wind of the form given in Eq. (1) up to a distance $r_\mathrm{w}$ from the explosion site, surrounded by a 'stagnation zone' of roughly constant density. Assuming the transition between the freely expanding wind and the stagnation zone is a strong shock front with a compression ratio of 4, and that the flow remains spherically symmetric, we have:

$$\rho(r) = \dot{M}/(\pi r_\mathrm{w}^2 v_\mathrm{w}) \text{ for } r > r_\mathrm{w} . \qquad (2)$$

The stagnation zone, in turn, is presumably confined within a much denser, slower wind which flowed from the progenitor during its red-giant phase of evolution, up to about 10,000 years ago. It is possible that the ring of dense material seen at about $r = 6 \times 10^{17}$ cm may form part of this wind (see McCray 1993 for a review).

A characteristic of all these calculations is that the shock front is strongly modified soon after cosmic ray injection is assumed to begin. This can be understood by considering the relationship between the expansion timescale and the timescale for the rise in cosmic ray pressure in the test-particle approximation (DBK). Nonlinear effects quickly become important and modify the shock once the pressure in cosmic rays becomes comparable to the ram-pressure of the medium entering the shock front. An equivalent viewpoint is that this saturation point is reached when the cosmic rays take a substantial fraction $\phi \lesssim 1$ of the kinetic energy of the matter flowing into the shock front from upstream. Writing $r_\mathrm{s}(t)$ for the radius and $v_\mathrm{s}(t)$ for the velocity of the modified shock front in the rest frame of the progenitor at time $t$, we can estimate the energy density in cosmic rays $E_\mathrm{C}$ as

$$E_\mathrm{C}(t) \approx \frac{\phi}{V(t)} \int_0^t \mathrm{d}t' \, 4\pi \left[r_\mathrm{s}(t')\right]^2 \frac{1}{2}\rho\left[r_\mathrm{s}(t')\right] \left[v_\mathrm{s}(t')\right]^3 , \qquad (3)$$

where $V(t)$ is the volume occupied by the cosmic rays, roughly equal to that of the shocked wind. We define a filling factor $f$ as follows

$$V(t) = f 4\pi \left[r_\mathrm{s}(t)\right]^3 / 3 . \qquad (4)$$

If $\rho_\mathrm{comp}$ is the overall compression ratio of the modified shock consisting of both sub-shock and precursor, then $1 > f > 1 - (1 - 1/\rho_\mathrm{comp})^3$.

An unsolved question in the theory of diffusive acceleration which is important for the determination of the efficiency $\phi$ is that of how particles are injected into the mechanism. The parameterisation of this process used in the two-fluid models of SN1987A is approximately the same as that used for models of supernova remnants. It is also similar to the more detailed kinetic models of supernova remnants presented by Berezhko et al. (1994). However, independent of the assumptions concerning injection, all treatments find a cosmic ray energy density in agreement with Eq. (3) once the shock is significantly modified.

The basic quantity required for a calculation of the gamma-ray production rate is the product of the target mass $M_\mathrm{target}$ and the energy density $E_\mathrm{C}$ of cosmic rays. The target mass is given by the mass in the shocked wind:

$$M_\mathrm{target}(t) = \int_0^t \mathrm{d}t' \, 4\pi \left[r_\mathrm{s}(t')\right]^2 \rho\left[r_\mathrm{s}(t')\right] v_\mathrm{s}(t') . \qquad (5)$$

During the free-expansion phase of the supernova, it is a good approximation to treat $v_\mathrm{s}$ as constant, in which case

$$E_\mathrm{C} M_\mathrm{target} = 3\phi M_\mathrm{target}^2 / (f 8\pi v_\mathrm{s} t^3) . \qquad (6)$$

Drury et al. (1994) have given a convenient method of estimating the gamma-ray production rate $\dot{N}_{\text{ph}}$ given the target mass and cosmic ray energy density:

$$\dot{N}_{\text{ph}} = qE_{\text{C}}M_{\text{target}}/\mu ,  \qquad (7)$$

where $\mu$ is the mean molecular weight of the progenitor wind and the quantity $q$ depends on the spectrum of the cosmic rays and the energy of gamma-rays considered. For a wind profile of the form $\rho \propto r^{-n}$, we have $M_{\text{target}} \propto t^{3-n}$ and $\dot{N}_{\text{ph}} \propto t^{3-2n}$. In the vicinity of the shock, the cosmic ray spectrum is expected to be relatively hard (Berezhko et al. 1994), and we therefore take $q = 10^{-17}$ for gamma-rays of energy greater than 1 TeV (Drury et al. 1994). Assuming a distance of 55 kpc, the flux from SN1987A whilst it was expanding into the region $r < r_{\text{w}}$ where the wind corresponded to a constant mass-loss rate is given by

$$F(>1\,\text{TeV}) = 2 \times 10^{-14} \frac{\phi}{f} \left(\frac{\dot{M}}{10^{-5}M_\odot\,\text{yr}^{-1}}\right)^2$$
$$\times \left(\frac{v_{\text{s}}}{2 \times 10^4\,\text{km s}^{-1}}\right) \left(\frac{v_{\text{w}}}{500\,\text{km s}^{-1}}\right)^{-2} \left(\frac{t}{1\,\text{yr}}\right)^{-1}$$
$$\text{photons s}^{-1}\,\text{cm}^{-2} . \qquad (8)$$

This is in rough agreement with the estimate for a galactic type II SN by Berezinskii & Ptuskin (1988), if we take into account the different wind speed and greater distance to SN1987A. According to the best-fit model of DBK, $r_{\text{w}} = 2 \times 10^{17}$ cm, and the shock entered the stagnation zone in 1990. This is in agreement with the increase in soft X-ray emission observed at about this epoch (Beuermann et al. 1994). It follows that at present the gamma-ray flux is given approximately by (Kirk et al. 1994b):

$$F(>1\,\text{TeV}) = 1.7 \times 10^{-13} \frac{\phi}{f} \left(\frac{\dot{M}}{10^{-5}M_\odot\,\text{yr}^{-1}}\right)^2$$
$$\times \left(\frac{v_{\text{s}}}{2 \times 10^4\,\text{km s}^{-1}}\right)^5 \left(\frac{v_{\text{w}}}{500\,\text{km s}^{-1}}\right)^{-2}$$
$$\times \left(\frac{r_{\text{w}}}{2 \times 10^{17}\,\text{cm}}\right)^{-4} \left(\frac{t}{8\,\text{yr}}\right)^3 \text{photons s}^{-1}\,\text{cm}^{-2} . \qquad (9)$$

This flux level is very sensitive to the shock speed, which we expect to decrease somewhat in the stagnation zone. It also depends on the assumption of spherical symmetry, which is certainly inappropriate at distances approaching that of the ring. Once the dense ring is encountered we may expect the target mass to increase rapidly. Assuming, conservatively, that the cosmic ray pressure remains at the level in the stagnation zone and that the entire 0.05 $M_\odot$ of the ring is engulfed, we find

$$F(>1\,\text{TeV}) = 3.3 \times 10^{-13} \frac{\phi}{f} \left(\frac{\rho(r > r_{\text{w}})}{10^{-22}\text{g cm}^{-3}}\right)$$
$$\times \left(\frac{v_{\text{s}}}{2 \times 10^4\,\text{km s}^{-1}}\right)^2 \text{photons s}^{-1}\,\text{cm}^{-2} . \qquad (10)$$

## 3. SN1993J and other RSNe

The radio emission of several supernovae has been successfully modelled as synchrotron radiation from relativistic electrons which move out with the shock front through the medium surrounding the progenitor (Chevalier 1982; Weiler et al. 1986). As the layer of absorbing material between the observer and the shock front gets thinner, the radio emission increases in strength, emerging first at high frequencies, and later at the more strongly absorbed low frequencies. The rising phase of the emission thus yields information about the density of the progenitor wind, and the speed at which the shock moves through it.

This approach does not include a theory of particle acceleration *per se*; it is simply assumed that a fixed fraction of the energy available at the shock front is converted into both magnetic field and relativistic electrons. The distribution function of the electrons is assumed to be a power-law. Detailed modelling of these sources using the diffusive acceleration mechanism is still in progress (Ball & Kirk 1995) but it is sufficient for the present to note that in every case the observations require a steeper power-law spectrum than is predicted by diffusive acceleration at a strong shock. This suggests that in every RSNe the shock front is modified by the pressure of cosmic rays, so that the energy density in these can be estimated from Eq. (3).

SN1993J is a particularly interesting RSN which is only 3.6 Mpc distant and appears to have been a late-type supergiant with a massive, slow stellar wind. According to Van Dyk et al. (1994), the time dependence of the free-free absorption of radio emission by the wind in front of the supernova shock indicates that the mass-loss rate $\dot{M}$ of the progenitor's wind was not constant, but rather decreased as the star evolved towards explosion. This conclusion is supported by the behaviour of the X-ray emission (Fransson et al. 1994). The density profile around the progenitor is believed to be $\rho \propto r^{-3/2}$, from which it follows that

$$M_{\text{target}}(r) = 2r\,\dot{M}(r)/(3v_{\text{w}}) , \qquad (11)$$

where $\dot{M}(r)$ refers to the mass-loss rate at a time $r/v_{\text{w}}$ before explosion. Van Dyk et al. (1994) find $\dot{M} = 1.5 \times 10^{-4} M_\odot\,\text{yr}^{-1}$ at $r = 4.1 \times 10^{16}$ cm, and thus the flux is

$$F(>1\,\text{TeV}) = 1.8 \times 10^{-12} \frac{\phi}{f} \left(\frac{\dot{M}(4.1 \times 10^{16}\,\text{cm})}{1.5 \times 10^{-4} M_\odot\,\text{yr}^{-1}}\right)^2$$
$$\times \left(\frac{v_{\text{s}}}{2 \times 10^4\,\text{km s}^{-1}}\right)^2 \left(\frac{v_{\text{w}}}{10\,\text{km s}^{-1}}\right)^{-2}$$
$$\text{photons s}^{-1}\,\text{cm}^{-2} , \qquad (12)$$

independent of the time of observation. Of course, the $\rho \propto r^{-3/2}$ density profile cannot extend indefinitely, since this would imply an infinitely large mass-loss. This suggests that the $t^{-1}$ behaviour of the constant $\dot{M}$ case (Eq. 8) should apply once the shock reaches a sufficiently large radius. The transition between these phases should also be evident from the radio data.

## 4. Discussion

From the point of view of particle acceleration theory, the most uncertain factors in the estimated gamma-ray fluxes are the efficiency $\phi$ and the filling factor $f$. The latter depends on the total compression ratio of the subshock plus precursor system and the extent to which the post-shock flow is decelerated. A total compression of 4 and constant rate of expansion leads to $f = 0.58$. However, the true value of $f$ may be significantly smaller, since not only do the cosmic rays soften the equation of state of the gas, they also remove energy from behind the shock by diffusion and both of these effects tend to increase $\rho_{\rm comp}$. The two-fluid simulations of SN1987A by DBK suggest typical overall compression ratios of $\rho_{\rm comp} \approx 10$, and have $f \approx 0.27$.

An estimate of the ratio $\phi/f$ can be obtained directly from simulations, since for an $r^{-n}$ density profile and constant shock speed it is simply related to the ratio of the cosmic ray pressure $P_{\rm C}$ to the ram pressure ahead of the shock front: $\phi/f = 2(3-n)(P_{\rm C}/\rho v_{\rm s}^2)$. In the models presented by DBK, this quantity rises during the free wind phase, and stabilises at a value of roughly $P_{\rm C}/\rho v_{\rm s}^2 \approx 0.5$, corresponding to an efficiency $\phi \approx 30\%$. It follows that the flux of gamma-rays of energy above 1 TeV from SN1987A, as predicted in Eq. (10), may become detectable, when the shock encounters the dense ring known to surround the supernova. Furthermore, Eq. (12) suggests that SN1993J may be detectable, since it is only a factor of five lower than the flux from the Crab Nebula – the best observed TeV source to date.

Such a high gamma-ray flux has direct implications for radiation in other bands. Electrons produced by the decay of $\pi^+$ mesons created in cosmic ray-nucleon interactions in SN1993J have a luminosity which can be estimated, to order of magnitude, as 50% of the gamma-ray luminosity: $L_{\rm e} \approx 2 \times 10^{39} {\rm erg\,s}^{-1}$. Such electrons cool predominantly by synchrotron radiation and produce luminosities in the X-ray bands of ROSAT, ASCA and OSSE which are typically a factor of a few below $L_{\rm e}$. This estimate is not too far below observed fluxes (Zimmermann et al. 1994; Tanaka et al. 1993; Leising et al. 1994). Clearly, an accurate calculation of the relationship between the yield of gamma-rays above 1 TeV and the expected X-ray flux would be very interesting. This, however, requires detailed knowledge of the cosmic ray spectrum, which is not provided by two-fluid simulations, but must await a kinetic treatment, such as that performed for the wind of SN1987A by Berezhko (1994).

Although we expect the two supernovae discussed above to be somewhat weaker than the brightest galactic SNRs (Aharonian et al. 1994) they have the advantage of being point sources for Čerenkov telescopes, and may thus be easier to observe. The detection of very high energy gamma-rays from either a supernova or a supernova remnant would provide strong support for the hypothesis that supernova remnants are responsible for accelerating galactic cosmic rays below $10^{14}$ eV, and would also be a valuable diagnostic of the acceleration mechanism.

*Acknowledgements.* We would like to thank A. Mastichiadis, E. Berezhko, F. Aharonian and H. Völk for helpful discussions.